\documentclass[conference]{IEEEtran}
\IEEEoverridecommandlockouts
\usepackage{cite}
\usepackage{amsmath,amssymb,amsfonts}
\usepackage{algorithmic}
\usepackage{graphicx}
\usepackage{textcomp}
\usepackage{xcolor}
\usepackage{fancyhdr}

\usepackage{booktabs}
\usepackage{multirow}
\usepackage{subfigure}

\def\BibTeX{{\rm B\kern-.05em{\sc i\kern-.025em b}\kern-.08em
    T\kern-.1667em\lower.7ex\hbox{E}\kern-.125emX}}
\begin{document}

\title{A Comprehensive Study of Accelerating IPv6 Deployment\\
}
\author{\IEEEauthorblockN{Tianyu Cui$^{1,2}$, Chang Liu$^{1,2}$,  Gaopeng Gou$^{1,2}$, Junzheng Shi$^{1,2}$, and Gang Xiong$^{1,2}$}
\IEEEauthorblockA{1.Institute of Information Engineering, Chinese Academy of Sciences\\
2.School of Cyber Security, University of Chinese Academy of Sciences\\
Beijing, China\\
 \{cuitianyu, liuchang, gougaopeng, shijunzheng, xionggang\}@iie.ac.cn}
}

\maketitle

\thispagestyle{fancy}
\fancyhead{}
\lhead{}
\lfoot{\textbf{\textsf{~~978-1-7281-1025-7/19/\$31.00~\copyright2019 IEEE}}}
\cfoot{}
\rfoot{}

\begin{abstract}
Since the lack of IPv6 network development, China is currently accelerating IPv6 deployment. In this scenario, traffic and network structure show a huge shift. However, due to the long-term prosperity, we are ignorant of the problems behind such outbreak of traffic and performance improvement events in accelerating deployment. IPv6 development in some regions will still face similar challenges in the future. To contribute to solving this problem, in this paper, we produce a new measurement framework and implement a 5-month passive measurement on the IPv6 network during the accelerating deployment in China. We combine 6 global-scale datasets to form the normal status of IPv6 network, which is against to the accelerating status formed by the passive traffic. Moreover, we compare with the traffic during World IPv6 Day 2011 and Launch 2012 to discuss the common nature of accelerating deployment.
Finally, the results indicate that the IPv6 accelerating deployment is often accompanied by an unbalanced network status. It exposes unresolved security issues including the challenge of user privacy and inappropriate access methods. According to the investigation, we point the future IPv6 development after accelerating deployment.

\end{abstract}

\begin{IEEEkeywords}
Internet, IPv6, Measurement, Big Data Analytics, Network Security 
\end{IEEEkeywords}

\section{Introduction}
IPv6 \cite{b1}  is the next-generation IP protocol at the network layer \cite{b2}. Due to the lack of IPv4 address space \cite{b4}, IPv6 has developed for a period of time to solve the problems of IPv4 network.  Nowadays, the development of IPv6 networks has reached a new stage. The quite large proportion of IPv6 traffic in the network should no longer be dismissed by researchers as an uninteresting rarity.  Researching what is going on in today's IPv6 world is essential to meet the increasing demand for traffic management required by future IPv6 networks.  


Since World IPv6 Day 2011\cite{b23} and Launch 2012 \cite{b24}, many countries in the world have been beginning to deploy IPv6 network equipment on a large scale, like China. By 2016, more than 220 countries and regions around the world have applied for IPv6 addresses. However, the use degree of IPv6 in various regions exist many differences, as of August 2017, IPv6 traffic penetration rate in China ranked 67th in the world, only 0.6\%  \cite{b34}. Under these circumstances,
At the end of 2017, China circulated an Action Plan for Promoting Scale Deployment of Internet Protocol Version 6 (IPv6) \cite{b22}. At present, China is vigorously promoting the deployment of IPv6 network equipment. During this major event, suddenly increased IPv6 traffic and the infancy of IPv6 networks could lead to unpredictable network status (we call it accelerating status in the paper). We conclude that regions in the future still have a possibility of accelerating development. Studying in the accelerating status will contribute to guiding future IPv6 development.

Our aim is to understand the accelerating status during the period of accelerating IPv6 scale deployment in China. 
A once-in-a-lifetime opportunity to observe technological change on such a grand scale, this is both practically and scientifically important. However, with a handful of exceptions, the previous body of measurement work has been invaluable at only a single aspect of IPv6 (e.g., traffic trends) and/or measuring on a confined area with limited traffic (e.g.,  one server or campus network). A broader approach combined with multiple means to observe the whole Chinese IPv6 network status is rarely been conducted. In addition, considerable work based on the basic IPv6 status. We have been accustomed to the peace and prosperity of IPv6 and lack of understanding of major world IPv6 events. In comparison to most of the work, our goal is to realize a big picture of the accelerating IPv6 status and discover the issues of the accelerating IPv6 network deployment.



In this paper, to achieve the goal, we conduct a comprehensive measurement of IPv6 nationwide network status during the period of accelerating IPv6 scale deployment in China. We implement a 5-month passive measurement from March to July in 2018. The work collects IPv6 traffic on a large IPv6 network that connects various IPv6 networks including the IPv6 backbone networks in China. For a better understanding of this specific period of IPv6 network, we assemble a set of passive collected IPv4 traffic and publicly-accessible datasets that speak to one or more aspects of IPv6 adoption to compare with our passive collected traffic. In addition to the traffic data, we actively scanned Alexa Top 10,000 global websites and Chinese websites in Alexa Top 1 million websites. According to the investigation, we conclude the nature of accelerating IPv6 deployment status and predict the future development trend.

\textbf{Our contributions can be summarized as follows:} 
\begin{itemize} 
\item We are the first one researching on the period of promoting large-scale IPv6 network deployment in China by using long-time passive traffic and comparing with passive collected IPv4 traffic, publicly-accessible datasets and the result of active scanning. This combination effectively unearths the real world of IPv6 and strongly confirms it.


\item We produce a measurement framework combining public datasets to form a normal IPv6 network status. It explores the conclusion of accelerating deployment by comparing normal status and accelerating status. 


\item We found that the accelerating IPv6 deployment is often accompanied by an unstable status. In addition, accelerating status is lack of security because of many issues including low content encryption rates and excessive IPv4/IPv6 transition usage.




\item We explore the common nature of the accelerating IPv6 deployment and give the prediction of future traffic trends. According to these, we provide advice for the future development of IPv6. The work will continue to operate to guide future network. 
\end{itemize}

The organizational structure for the rest of this paper is as follows. In section \ref{P2}, we review the previous studies. Section \ref{P3} explains the details of our measurement methodology and data collection. We depict the real-world IPv6 network status and analyze the IPv6 accelerating status in Section \ref{P10} by tracing the address distribution, traffic trends, services deployment, and communication protocols. In section \ref{P12}, we talk about security issues and common nature during the accelerating of large-scale deployment. The discussion about the current accelerating status in China and future IPv6 development are shown in Section \ref{P5}. 

\section{Related Work}\label{P2}

In the field of traffic measurement, there are many papers in the literature that offer valuable data on the IPv6 adoption process from various perspectives. For instance, Several studies characterize IPv6 traffic from the perspective of one or more ISPs (e.g.,  \cite{b13}), Other work examines IPv6 in one single aspect, such as DNS (e.g., \cite{b12}). In contrast to much of these studies, we only focus on the whole status of IPv6 networks and the nature of the traffic during the accelerating deployments.



However, few studies investigate a wide range of comprehensive measurements on the IPv6 network environment in a certain country or region. In order to solve this problem, Czyz et al. \cite{b25} explored twelve metrics using ten global-scale datasets to create a broad measurement of IPv6 adoption. Moreover, they collected unclaimed traffic on the IPv6 Internet by announcing five large /12 covering prefixes and analyzed the nature of this traffic across regions \cite{b26}. However, they focus on the overall ecosystem rather than fine-grained points. Different from their study, our work is performed during the accelerating deployments in China and we combine passive and active measurements to reveal the potential issues during the major IPv6 event.

One of the closest to our work in substance is Sarrar et al. \cite{b27}. They explicitly explore the IPv6 deployment event during IPv6 World Day on June 8, 2011. In comparison, our work not only focuses the change point of the IPv6 traffic during the event, but we also describe the whole IPv6 networks status and explore the common issues during the period of accelerating deployment. In addition, we compare with the datasets captured during these major world events to discover the common nature of the change.

\begin{table*}[htbp]
\caption{Dataset summary showing the time period, scale, and collection method of the datasets we analyzed.}
\begin{center}

\begin{tabular}{|l|c|l|c|c|}
\hline
\textbf{Dataset}&\textbf{Time Peroid}&\textbf{Scale}&\textbf{Category}&\textbf{Collection}\\
\hline
MAXMIND GeoLite2 City&10 July 2018&global Geo-IP dataset&\multirow{6}*{\shortstack{N}}&\multirow{7}*{\shortstack{Public}}\\
\cline{1-3}
RIR Address Allocations&20 March - 10 July in 2018&565 daily allocation snapshots&&\\
\cline{1-3}
Routing: Route Views&20 March - 10 July in 2018&113 BGP table snapshots&&\\
\cline{1-3}
Google IPv6 Client Adoption&20 March - 10 July in 2018&daily global samples&&\\
\cline{1-3}
Verisign TLD Zone Files&20 March - 10 July in 2018&daily snapshots of A and AAAA records (.com \& .net)&&\\
\cline{1-3}
Alexa Top Sites&10 July 2018&Top 10K global sites, 1,696 Chinese sites in Top 1M&&\\
\cline{1-4}

CAIDA IPv6 Day and Launch Day&8 June 2011 and 6 June 2012&1TB pcap files of anonymized passive traffic&A&\\
\hline  
\hline  

IPv6 Real Traffic Dataset&20 March - 10 July in 2018&170 million flows captured in 5 months&A&Passive\\
\cline{1-4}
IPv4 Real Traffic Dataset&20 March - 30 April in 2018&35 million flows captured in 1 month&N&Measurement\\
\hline
\multicolumn{5}{l}{$^{\mathrm{*}}$Category N and A are the datasets respectively used for building normal status and accelerating status.}\\

\end{tabular}
\label{table1}
\end{center}
\end{table*}

\section{Methodology and Dataset}\label{P3}

In this part, we illustrate our passive and active measurement methodology and show all the datasets used in this paper.

\begin{figure}[htbp]
\begin{center}
\scalebox{0.29}{
\centerline{\includegraphics{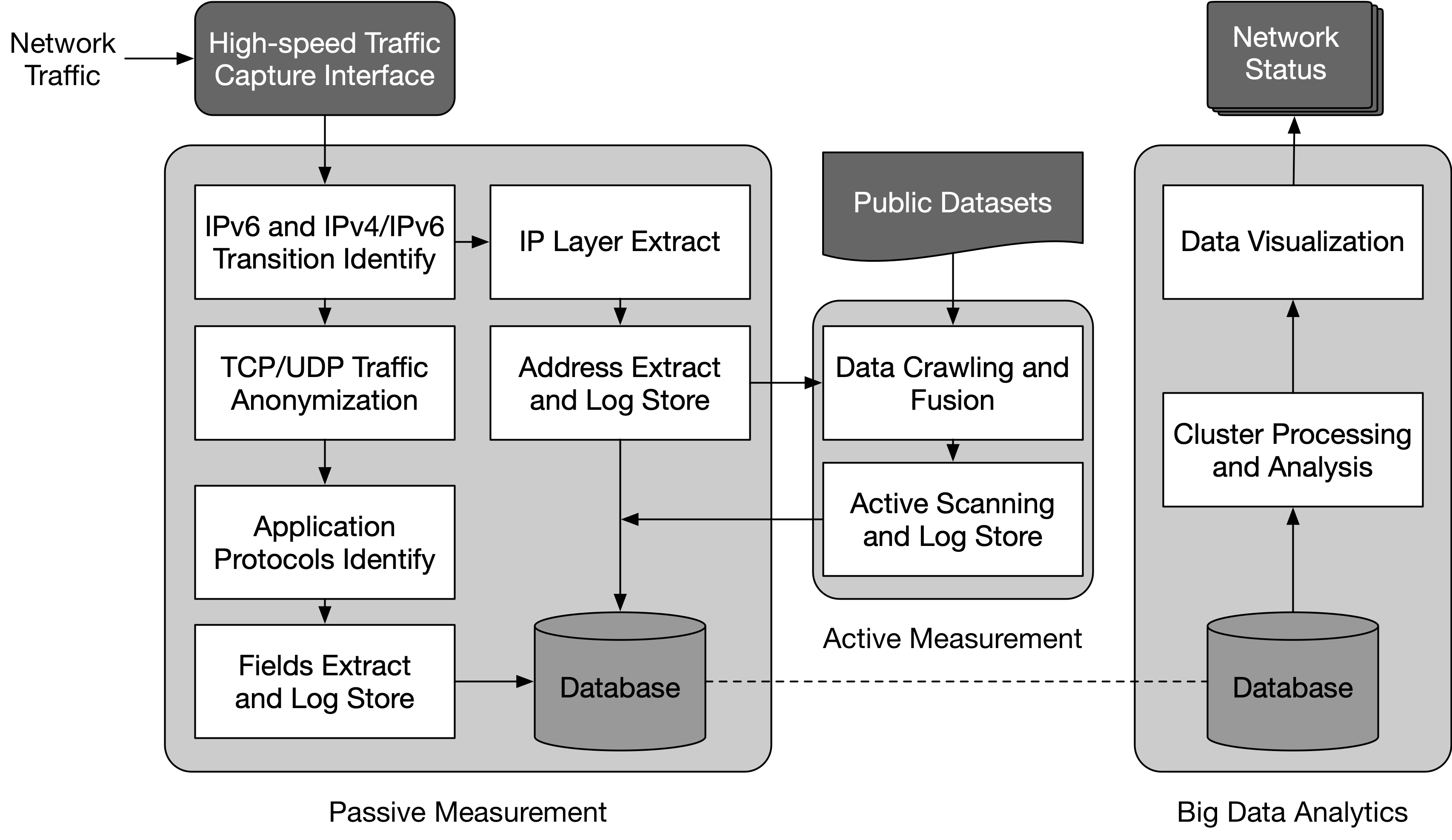}}
}
\caption{Measurement Framework.}
\label{framework}
\end{center}
\end{figure}


\begin{figure*}[htbp]
\centering
\subfigure[World]{       
\label{1}
\begin{minipage}[t]{0.31\linewidth}
\centering
\includegraphics[width=5.8cm]{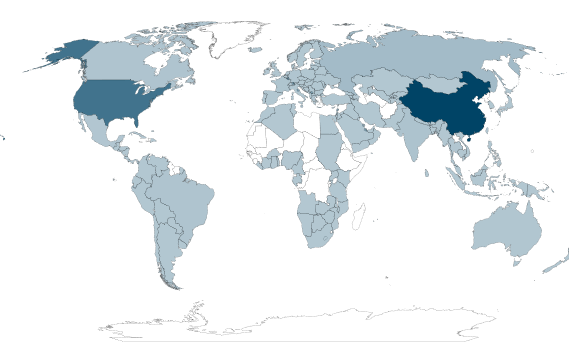}
\end{minipage}
}
\subfigure[China]{ 
\label{2}
\begin{minipage}[t]{0.31\linewidth}
\centering

\includegraphics[width=3.5cm]{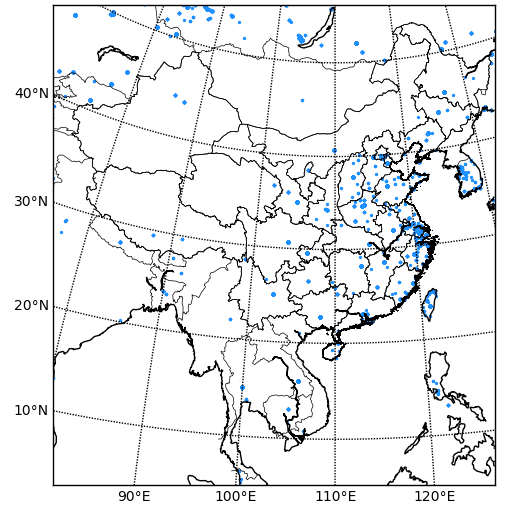}
\end{minipage}
}
\subfigure[the United States]{ 
\label{3}
\begin{minipage}[t]{0.31\linewidth}
\centering
\includegraphics[width=4.8cm]{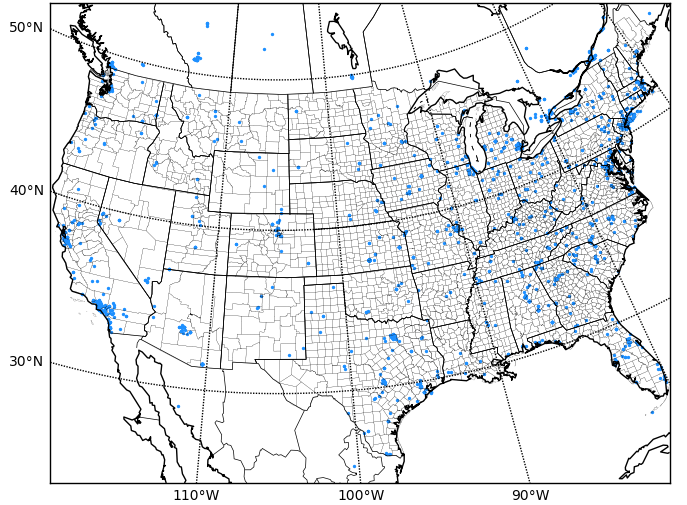}
\end{minipage}
}
\centering
\caption{IPv6 address distribution in the World, China and the United States.}
\label{fig1}
\end{figure*}

\subsection{Measurement Methodology}

In this paper, we implement a large-scale passive measurement on an IPv6 network from March to July in 2018. We choose the traffic during this period because it is in the early stage of deployment. Due to the network status tends to be stable in the middle and late stages of deployment, early-stage traffic is more likely to reflect the comprehensive impact of accelerating deployment. Our measurement is continuously working on the network named \textit{China Unicom}, which is provided by one of the three largest communication operators in China. It connects China's various IPv6 networks including \textit{CERNET} and \textit{CERNET2}. Our measurement framework is shown in Figure \ref{framework}. 
\begin{itemize}
\item Passive measurement. In the passive measurement, the program first identifies IPv6 traffic (including native IPv6 and IPv6 tunnel traffic) by address format and protocol stack structure. It separates the network layer and the transport layer for anonymization. We respectively extract the active IPv6 address and application layer protocol fields into the database. 

\item Active measurement. The program automatically crawls daily public datasets for analysis or scanning. After data preprocessing, we actively scan the Alexa Top sites \cite{b31} and active IPv6 address by sending the IPv6 TCP SYN to all the remote ports or using automatic jwhois query \cite{b37}. After receiving the response, we record these result and store into the database.

\item Big data analytics. After data storage, we use Spark \cite{b38} for fast big data analysis. The program can selectively analyze the current state of the network from one or more standards. Finally, data visualization is used to obtain an observation of the overall state of the network.
\end{itemize}

For ethical considerations, in the passive measurement process, we only extract the key fields for statistical analysis. Our measurement emphasizes the whole network status instead of personal privacy information. For active measurement, we only conduct scientific research and perform active detection processes in accordance with standardized measurement methods. 

\subsection{Dataset}


Table \ref{table1} summarizes all the datasets we analyzed in this paper. We utilize public datasets to form the normal network status and compare normal status  (Category N) and accelerating status  (Category A) to describe the accelerating IPv6 deployment. Among them, IPv4 real traffic dataset is captured by the same methodology as the IPv6 real traffic dataset.

\section{IPv6 Accelerating Status During the Accelerating of Large-scale Deployment in China}\label{P10}

In our measurement, we use the collected dataset to display the real world of the accelerating IPv6 networks during the large-scale deployment in China. For describing IPv6 networks in the period, we conduct an in-depth analysis of four points of view: address distribution, traffic trend, services deployment, and protocols in IPv6 networks.

\begin{table}
\caption{Usage of IPv6 Networks in China}
\begin{center}
\begin{tabular}{|c|c|c|c|}
\hline
\textbf{Net Name}&\textbf{Description}&\textbf{ASN}&\textbf{Addresses}\\
\hline
\hline
CU-CN&Unicom&4837&1,257,820\\
\hline
CNGI-CERNET2&CERNET2&23910&145,126\\
\hline
CERNET-CN&CERNET Backbone&23910&48,086\\
\hline
RELIANCEJIO-IN&Reliance Jio$^{\mathrm{1}}$&55836&38,654\\
\hline
CLOUDFLARE&Cloudflare, Inc.&-&26,275\\
\hline
PKU6-CERNET2&Peking University&23910&12,495\\
\hline
CERNET2-TSH6$^{\mathrm{2}}$&Tsinghua University&45576&8,742\\
\hline
ZSU6-CERNET&Zhongshan University&4538&8,165\\
\hline
ZZU6-CERNET2&Zhenzhou University&-&8,117\\
\hline
TELSTRAINTERNET$^{\mathrm{3}}$&Telstra Internet&-&7,605\\
\hline

\multicolumn{4}{l}{$^{\mathrm{1}}$Reliance Jio means Reliance Jio Infocomm.}\\
\multicolumn{4}{l}{$^{\mathrm{2}}$CERNET2-TSH6 means TCERNET2-TSINGHUA6}\\
\multicolumn{4}{l}{$^{\mathrm{3}}$TELSTRAINTERNET means TELSTRAINTERNET41-AU-20041202}

\end{tabular}
\label{table2}
\end{center}
\end{table}

\subsection{Address Distribution}

During the geolocation measurement process, we explore the geographical distribution of all active IPv6 addresses that appear in the network. The Geo-IP database we used is provided by MAXMIND \cite{b33}. Figure \ref{fig1} reveals the distribution status of IPv6 addresses in the World, China, and the United States. As the samples are collected on the vantage point in China, the results mean the distribution of the world's active IPv6 addresses that communicate frequently with China. At the present stage, Chinese IPv6 users rarely have access to many areas, like all over Africa. In addition, we show a detailed distribution in China and the United States. During the period of accelerating deployment,  it is obvious that Chinese IPv6 distribution is more concentrated and unbalanced. The deployment of future network structure will start from the east to the west.

\begin{figure}
\begin{center}
\scalebox{0.3}{
\centerline{\includegraphics{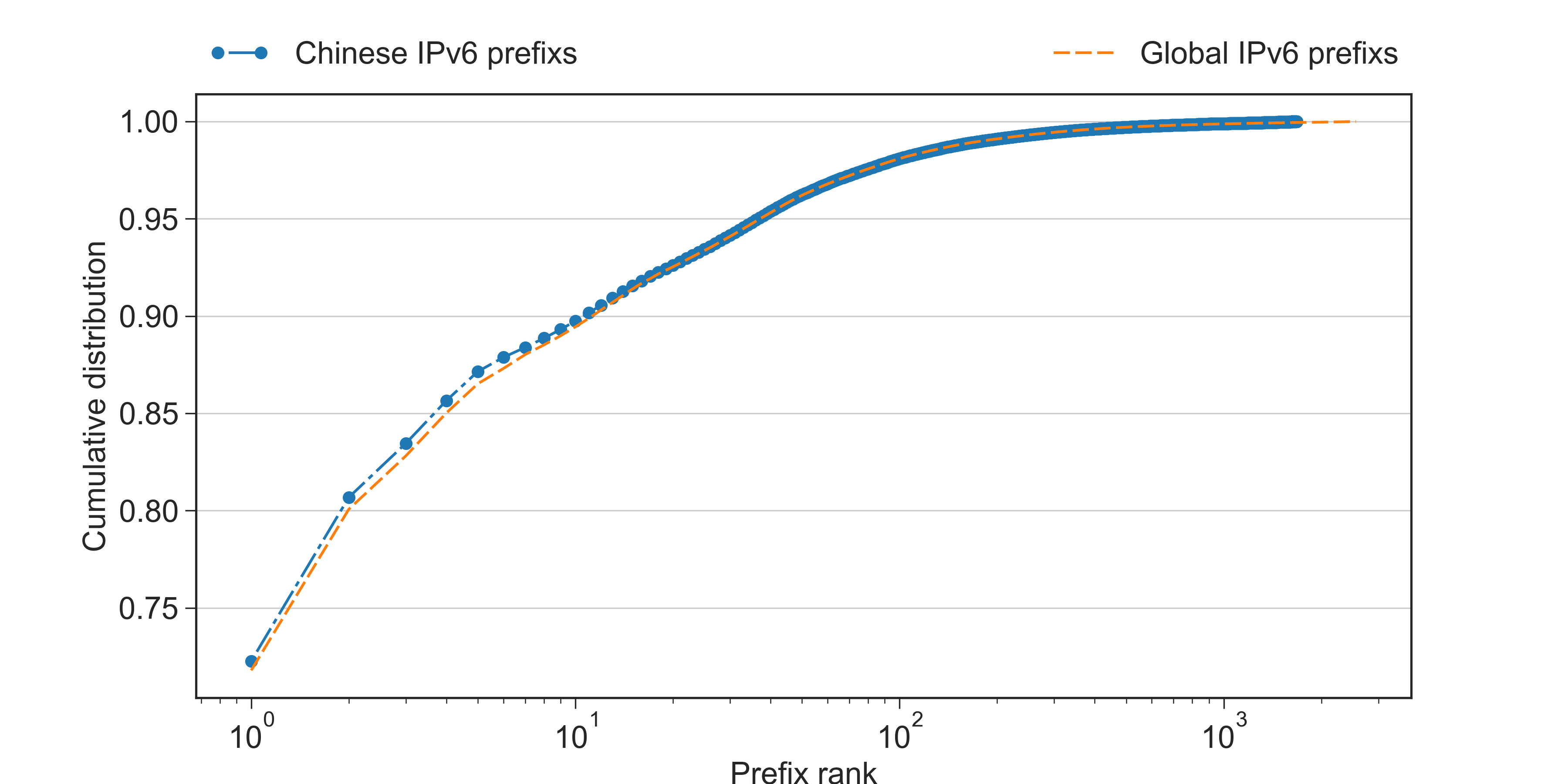}}
}
\caption{Distribution of IPv6 AS prefix rank.}
\label{fig2}
\end{center}
\end{figure}

For the accelerating status of network deployment in China, we have statistics on the specific distribution of IPv6 addresses in Table \ref{table2}. In the measurement, nearly two-thirds of the active IPv6 addresses belong to \textit{China Unicom} networks, named \textit {CU-CN}. As the most essential IPv6 networks in China, \textit{CERNET} and \textit{CERNET2} occupy a large quantity. The rest of The IPv6 hosts usually come from the Chinese University that connects with \textit{CERNET} and \textit{CERNET2} such as \textit{Peking University} and \textit{Tsinghua University}. The cumulative distribution of prefix entities ranked by amount is described In Figure \ref{fig2}. We discover that unique addresses in the traffic respectively fall into 1,680 and 2,558 prefixes of China and the whole. In the prefix level, top 10 prefixes dominate approximately 90\% of unique IPv6 addresses. This shows IPv6 hosts are more concentrated on several main AS prefixes, the monotonous IPv6 network lacks various hosts participate.

\begin{figure}
\begin{center}
\scalebox{0.24}{
\centerline{\includegraphics{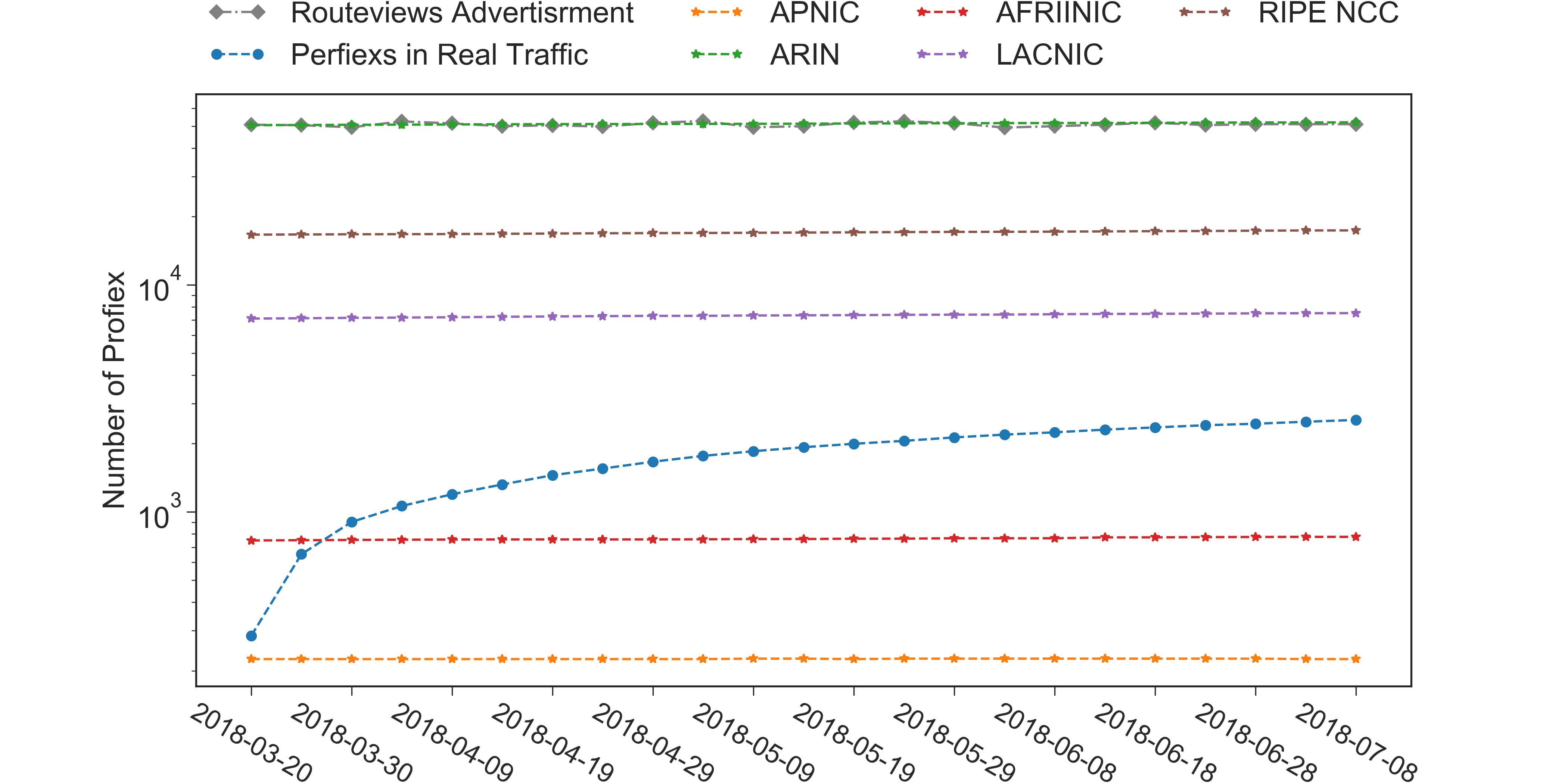}}
}
\caption{The world allocated, advertised and Chinese real-traffic prefixes.}
\label{fig3}
\end{center}
\end{figure}

In order to explore the usage and distribution of IPv6 prefixes,  we collect the world allocated and advertised prefixes compared with the real traffic. The allocation data is collected from the daily snapshot of the blocks of IPv6 addresses in five regional Internet registries (RIRs) \cite{b28}. In addition, we obtain the global routing table snapshots from the Route Views project \cite{b29} and extract the number of prefixes announced on the first 15 minutes of each day.  Figure \ref{fig3} shows the number of the world allocated, advertised and Chinese real-traffic prefixes over time. We know that calculating the prefixes in passive traffic requires long-term accumulation. Except for the statistics on early days, we find 858 IPv6 prefixes on April 1, 2018. On July 10, 2018, 2,559 IPv6 prefixes were collected—an increase of 3-fold over the course of three months. Even at the end of our measurement, the number of IPv6 prefixes increases by 10.1 per day, while the world allocation growth per day is 0.1 in APNIC, 5.2 in ARIN, 5.3 in RIPE NCC, 2.3 in LACNIC, 0.2 in AFRINIC. The number of advertised prefixes also not have obvious growth. This shows the rapid growth of IPv6 in China from a global perspective.

%

\begin{figure}
\begin{center}
\scalebox{0.35}{
\centerline{\includegraphics{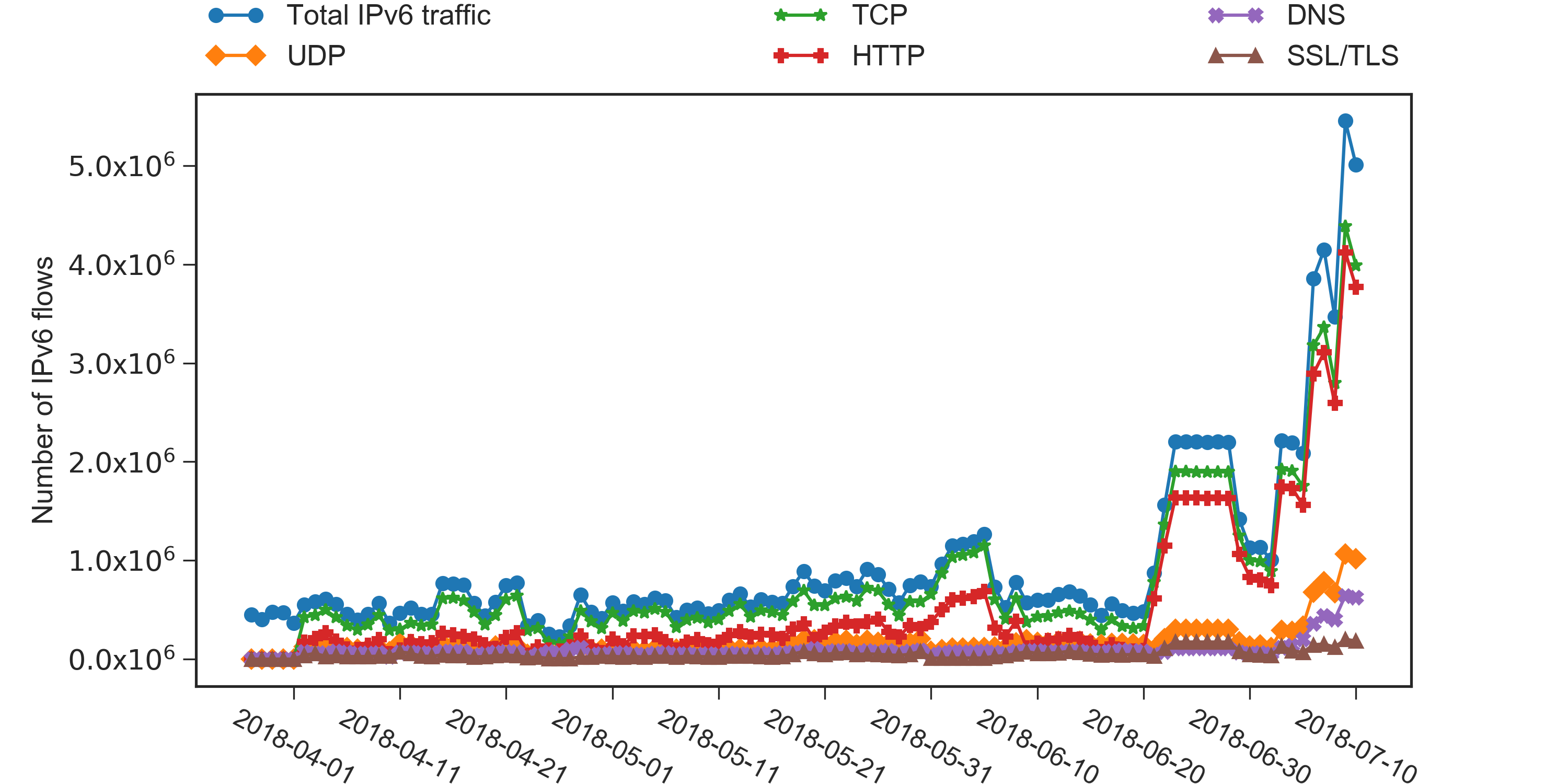}}
}
\caption{The trend of IPv6 flows from April to July 2018.}
\label{fig4}
\end{center}
\end{figure}

\begin{figure}
\begin{center}
\scalebox{0.47}{
\centerline{\includegraphics{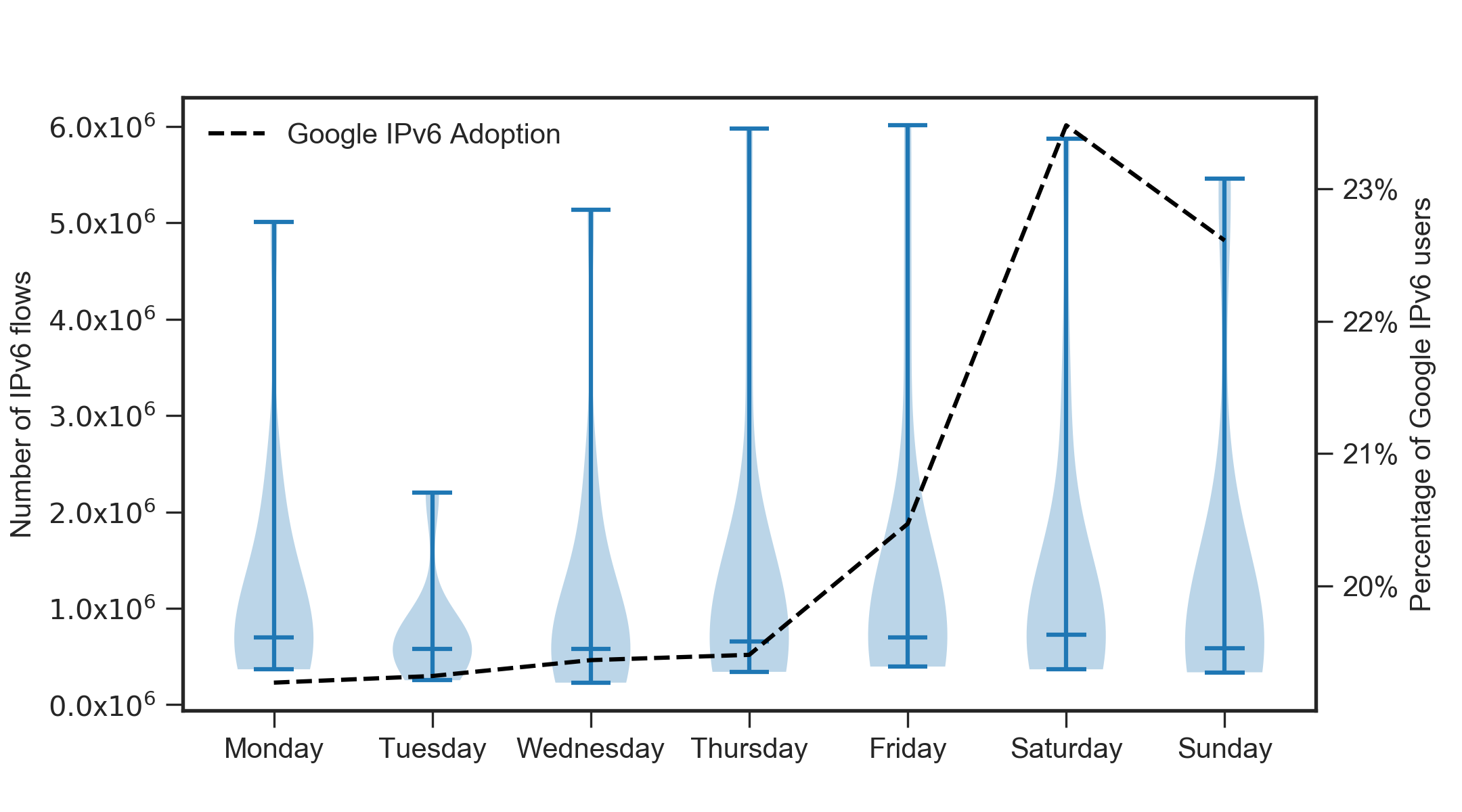}}
}
\caption{Google IPv6 adoption and Chinese IPv6 traffic in a week.}
\label{fig5}
\end{center}
\end{figure}

\subsection{Traffic Trends}

The IPv6 traffic trend is another observation strategy in our measurement. We continuously observed the traffic changes from March to July in 2018 and recorded the trend by our measurement framework. Figure \ref{fig4} shows the detail of the traffic change. Due to the accelerating IPv6 deployment and 4G networks are enabled to support IPv6 in China, IPv6 traffic has increased dramatically in the short term. The traffic volume has an increase of 10-fold. We infer that the great change event happened during June to July 2018 because of the Dramatic changes in traffic during that time. In addition, to determine whether more private or more office connections use IPv6, Figure \ref{fig5} reveals the difference in IPv6 traffic in a week. The difference of IPv6 traffic user usage on the first three days in a week is obviously less than the other days in a week, especially on Tuesday. The use of IPv6 traffic reaches the highest average value on Saturday. It indicates that IPv6 users more likely to use IPv6 networks on Monday, Friday, and Saturday because of the high average number of IPv6 flows. Moreover, Google \cite{b30} continuously collects statistical information about the adoption of IPv6 on the Internet. In Figure \ref{fig5}, Google IPv6 user usually access websites on the weekend. The IPv6 usage trend of Google users in a week is quite different from the IPv6 traffic during the deployment. We infer that at the present stage, Chinese IPv6 traffic usage mode is different from traditional user pattern. The accelerating network is made up of many complicated factors which are not controlled by users. 

\subsection{Service Deployment}\label{test}

\begin{figure}
\begin{center}
\scalebox{0.45}{
\centerline{\includegraphics{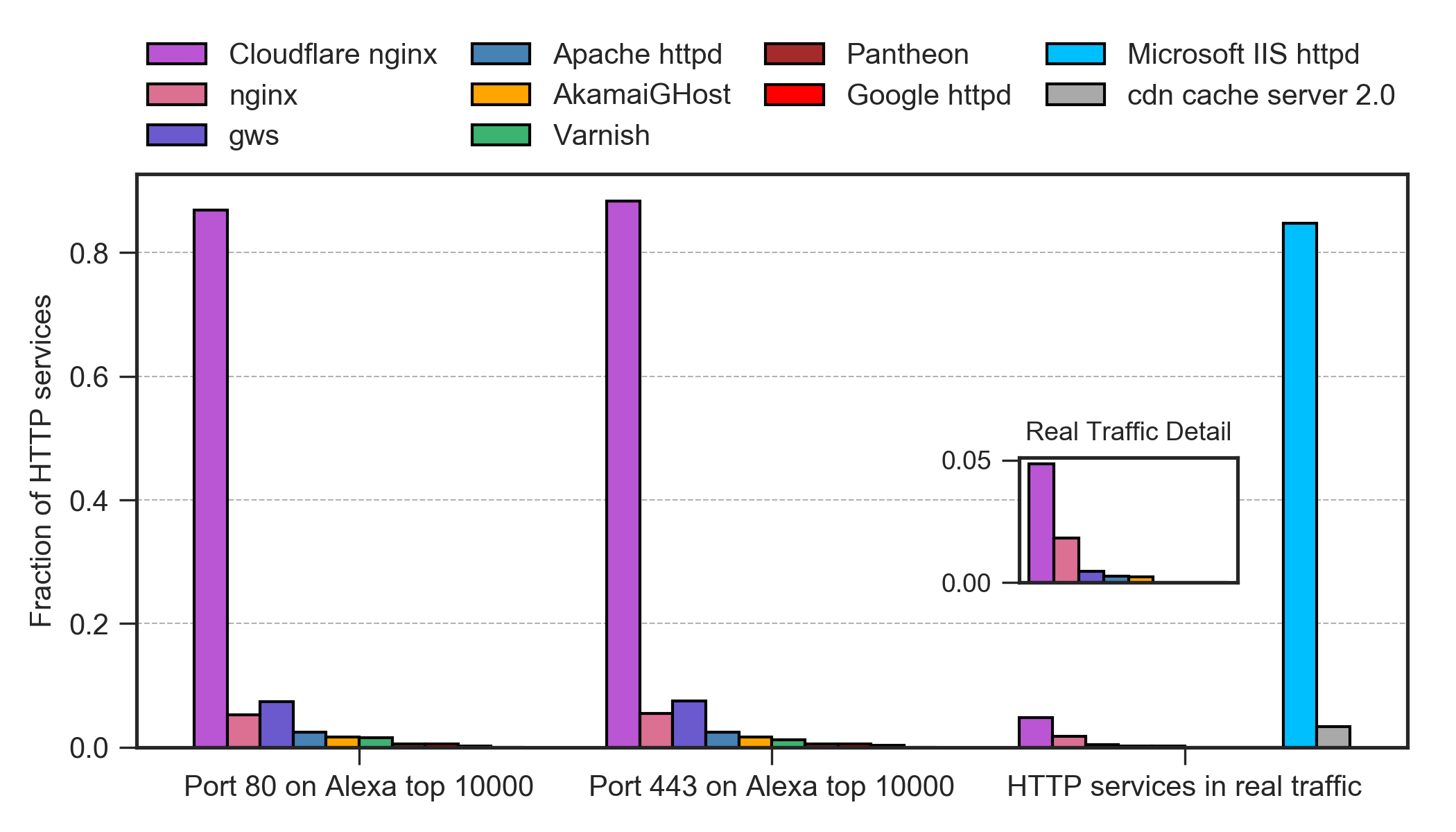}}
}
\caption{Deployment of server software on IPv6 websites.}
\label{fig6}
\end{center}
\end{figure}

\begin{figure}
\begin{center}
\scalebox{0.45}{
\centerline{\includegraphics{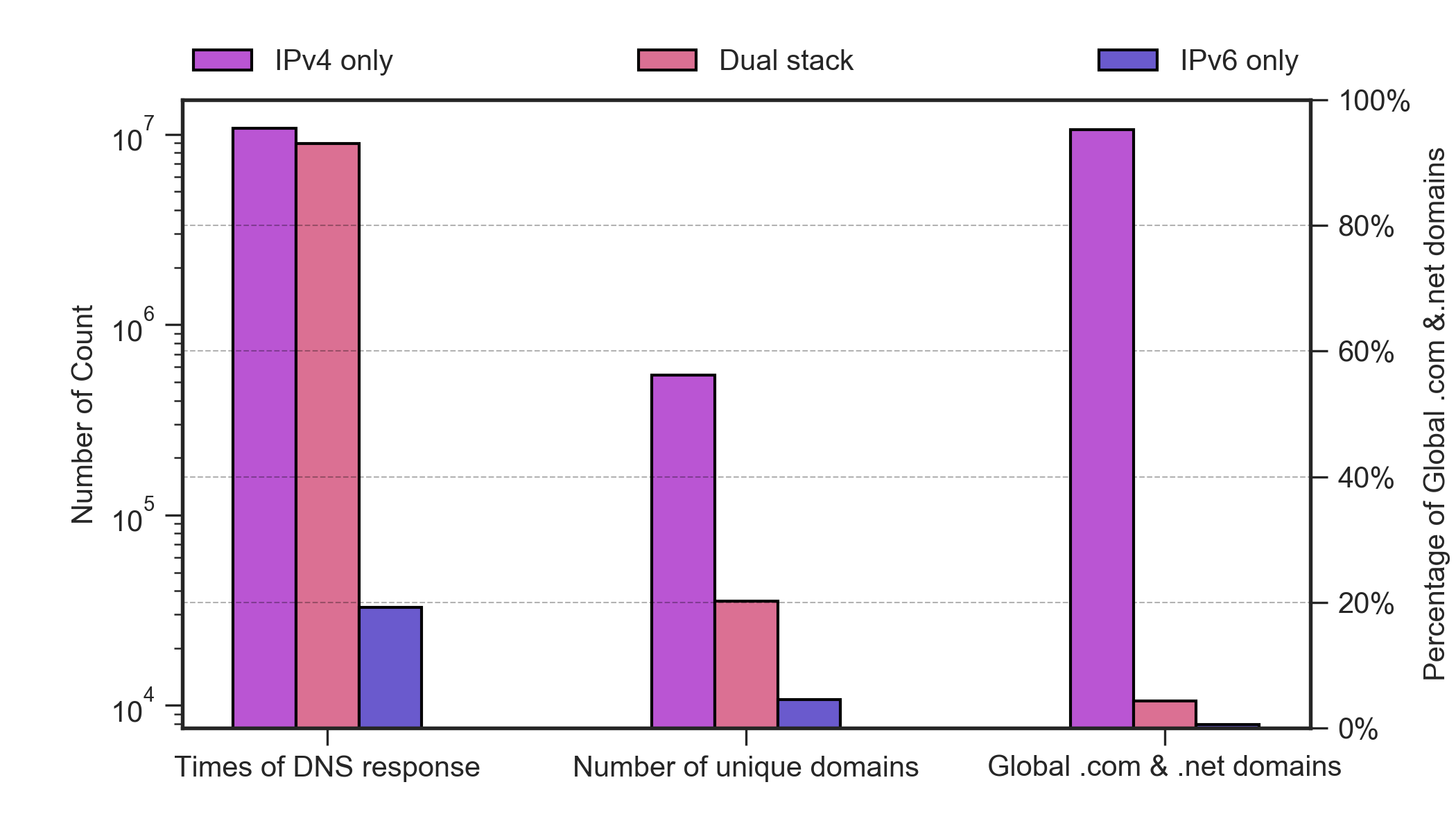}}
}
\caption{IPv4-only, IPv6-only and dual-stack domains.}
\label{fig7}
\end{center}
\end{figure}

For the servers,  the top 10 DNS answer domains, SSL/TLS SNI (Server Name Indication) and HTTP host are listed in Table \ref{table3}. Large Internet operators like \textit{Google}, \textit{Microsoft}, and \textit{Facebook} appear in the list as usual. Among them, \textit{Microsoft} occupies the most traffic. In the SSL/TLS SNI list, a porn website domain ranked first, which proves that many IPv6 users communicate with hidden or illegal sites. In addition, we actively scanned the Alexa Top 10,000 global websites and 1,696 Chinese domains in Alexa Top 1 million domains. Table \ref{table4} lists the service rankings they have opened. However, only 2,161 and 9 websites in the two sets respond to our IPv6 scanning. The proportion is 21.69\% and 0.005\% respectively, which shows an extreme scarce of Chinese IPv6 websites deployment. Except for the HTTP related services, a few sites provide SSH, FTP, and other services without filtering.

\begin{table}
  \caption{Top 10 Domains in three Protocols}
  \begin{center}
  \label{table3}
  \begin{tabular}{|c|c|c|c|}
   \hline
    \textbf{No}&\textbf{DNS Answer}&\textbf{SSL/TLS SNI}&\textbf{HTTP Host}\\
    \hline
    \hline
   1&*.msftncsi.com&141tube.com&*.windowsupdate.com\\
   \hline
   2&*.cnxt.com&*.google.com&*.microsoft.com\\
   \hline
   3&*.digitalturbine.com&*.adobe.com&*.msn.com\\
   \hline
   4&*.facebook.com&tu2.ttt669.com&bc.postcc.us\\
   \hline
   5&*.amazonaws.com&*.facebook.com&xxx55tp.com\\
   \hline
   6&iherbtest.com&*.microsoft.com&xinhuanet.com\\
   \hline
   7&epupdate.hmapi.com&768ii.com&google.com.hk\\
   \hline
   8&gwa.fe.bosch.de&highwebmedia.com&msftncsi.com\\
   \hline
   9&*.aws.af.cm&icloud.com&exoclick.com\\
   \hline
   10&google.com&exoclick.com&verisign.com\\
   \hline
\end{tabular}
\end{center}
\end{table}

\begin{table}
  \caption{Services Opened in the IPv6-enabled Sites}
  \begin{center}
  \label{table4}
  \begin{tabular}{|c|c|c|c|c|c|}
    
    \hline
   \multicolumn{3}{|c|}{\multirow{3}*{\textbf{\shortstack{Alexa Top 10,000\\Global Sites}}}}&\multicolumn{3}{|c|}{\multirow{3}*{\textbf{\shortstack{Chinese Sites in Alexa Top \\1 Million Global Sites}}}} \\
    
    \multicolumn{3}{|c|}{~}&\multicolumn{3}{|c|}{~}\\
    \multicolumn{3}{|c|}{~}&\multicolumn{3}{|c|}{~}\\
    \hline
    \textbf{Service}&\textbf{Sites}&\textbf{Percent}&\textbf{Service}&\textbf{Sites}&\textbf{Percent}\\ 
    \hline
    http/80&1,903&19.03\%&http/80&9&0.005\%\\
    \hline
    https/443&1,872&18.72\%&https/443&8&0.004\%\\
    \hline
    http/8080&1,598&15.98\%&http/8080&5&0.003\%\\
    \hline
    httpAlt/8443&1,590&15.90\%&httpAlt/8443&5&0.003\%\\
    \hline
    ssh/22&60&0.006\%&ftp/21&2&0.001\%\\ 
    \hline
    \hline
    Totally&10,000&100\%&Totally&1,696&100\%\\
    \hline
\end{tabular}
\end{center}
\end{table}

\begin{figure*}
\centering

\subfigure[SSL/TLS version]{   
\begin{minipage}[t]{0.31\linewidth}
\centering
\includegraphics[width=6cm]{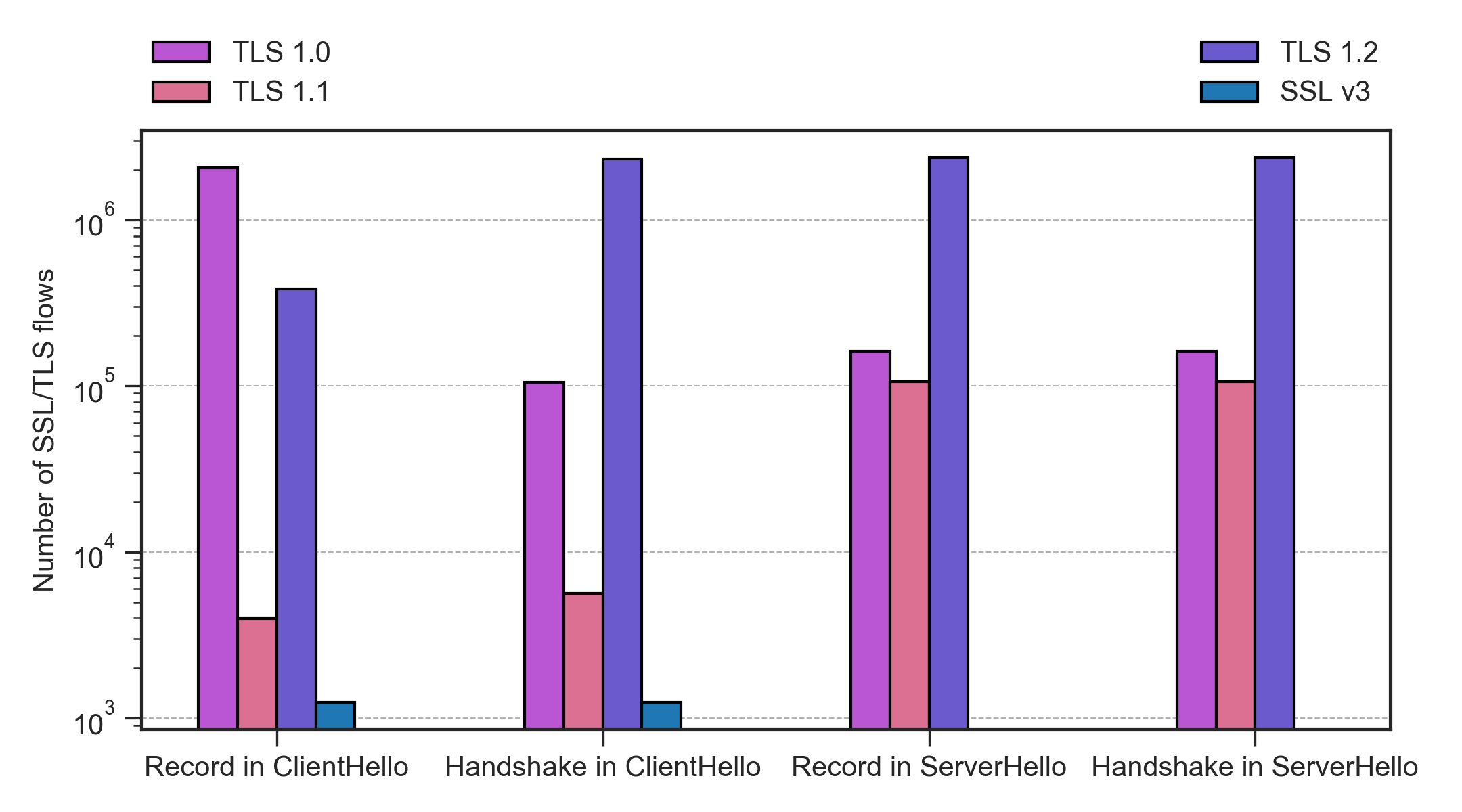}
\end{minipage}%
}
\subfigure[Certificate issuer]{   
\begin{minipage}[t]{0.31\linewidth}
\centering
\includegraphics[width=6cm]{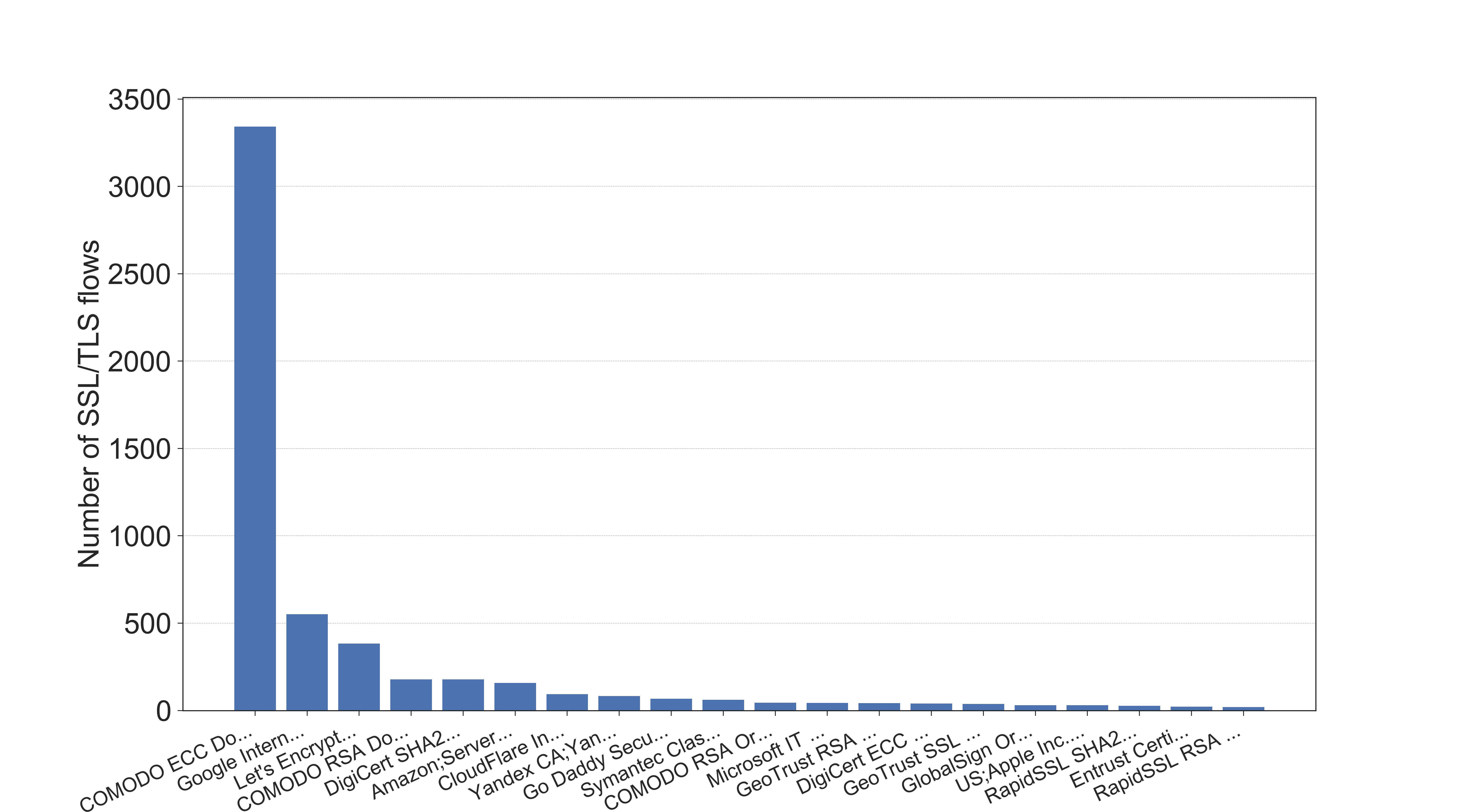}
\end{minipage}%
}
\subfigure[Certificate validity]{   
\begin{minipage}[t]{0.31\linewidth}
\centering
\includegraphics[width=6cm]{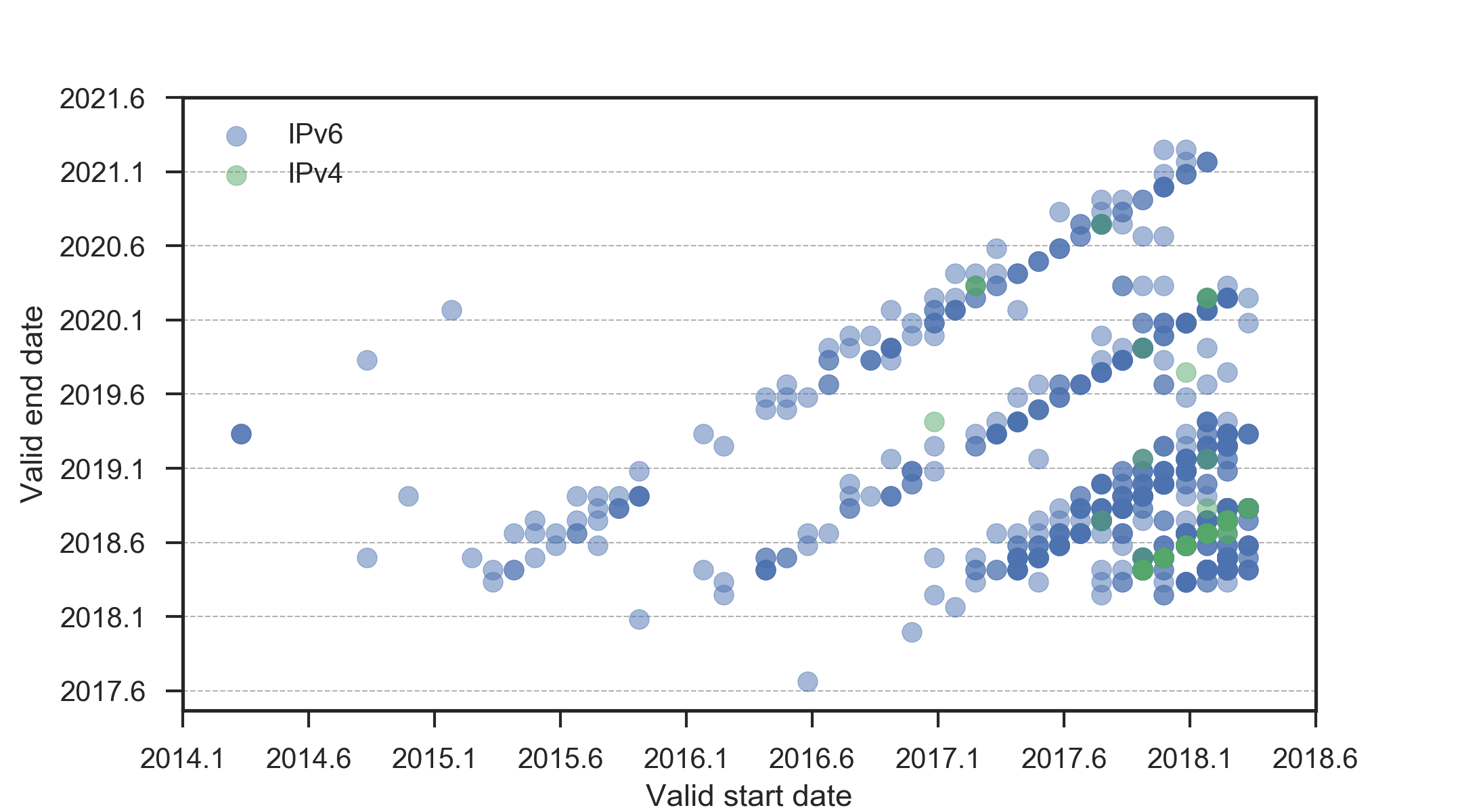}
\end{minipage}
}
\centering
\caption{SSL/TLS version, certificate issuer and valid time in SSL/TLS flows.}
\label{fig8}
\end{figure*}

\begin{figure*}
\centering
\subfigure[Content type field]{   
\begin{minipage}[t]{0.32\linewidth}
\centering
\includegraphics[width=6cm]{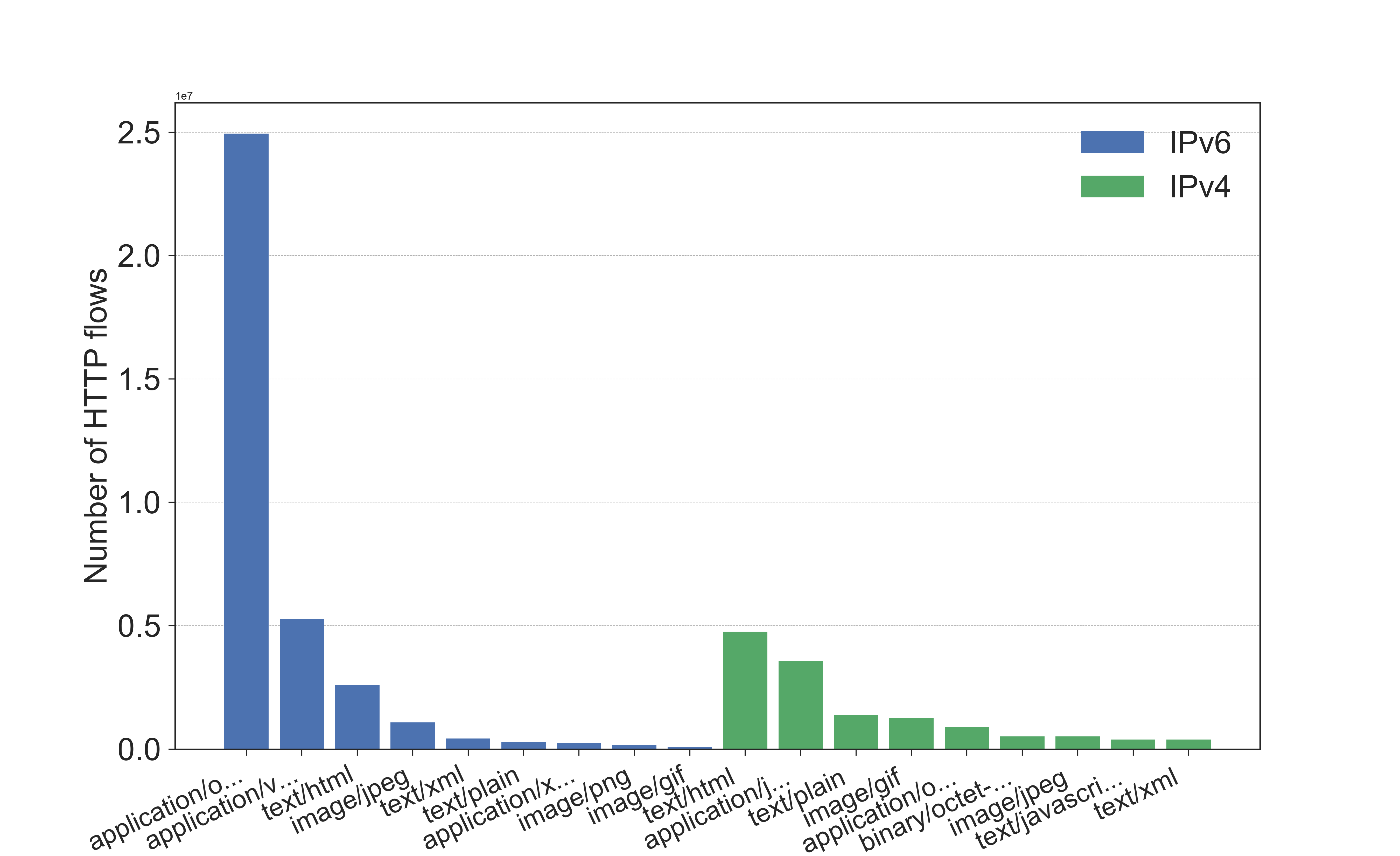}
\end{minipage}%
}
\subfigure[Response line field]{   
\begin{minipage}[t]{0.32\linewidth}
\centering
\includegraphics[width=6cm]{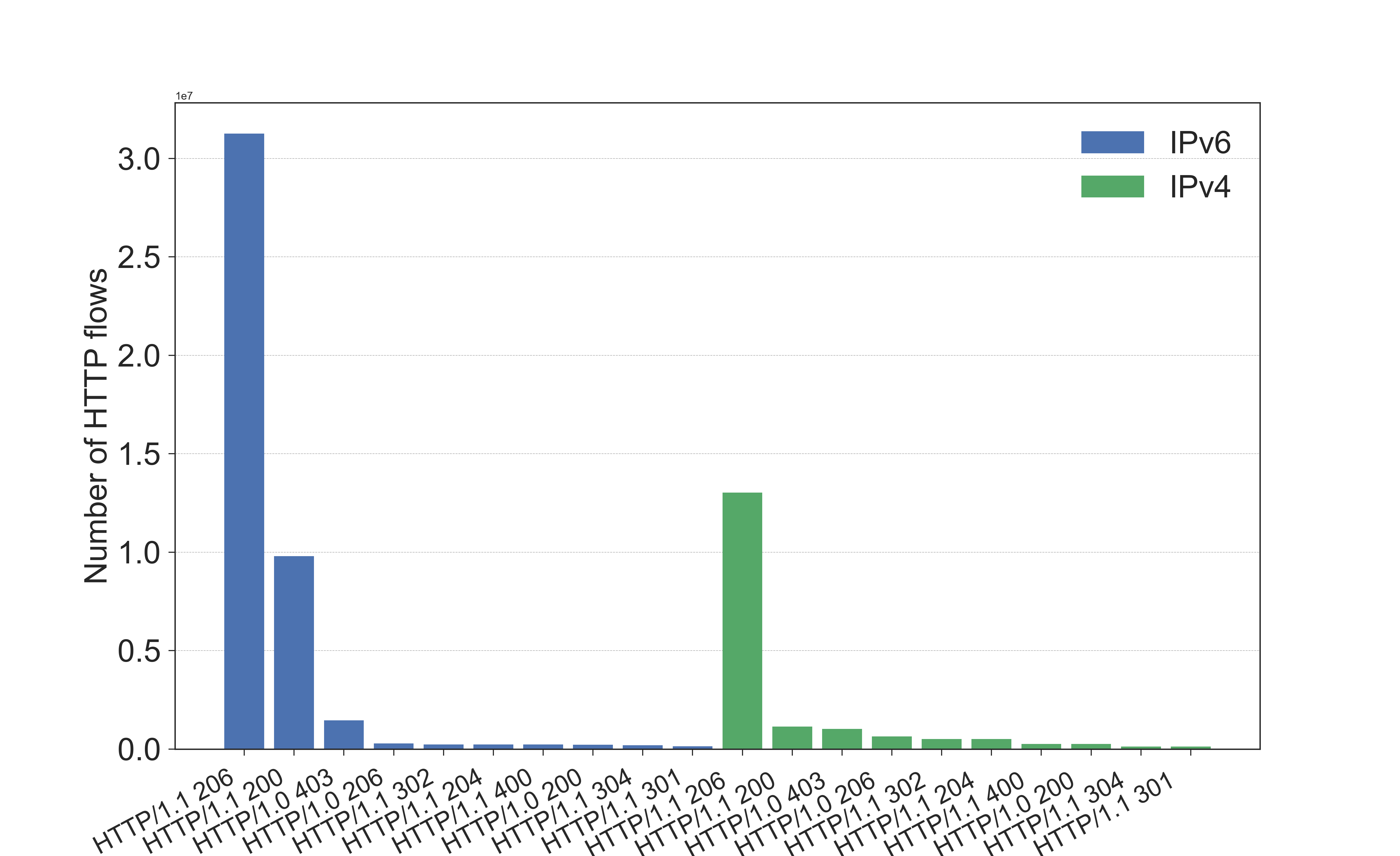}
\end{minipage}%
}
\subfigure[Via field]{   
\begin{minipage}[t]{0.32\linewidth}
\centering
\includegraphics[width=6cm]{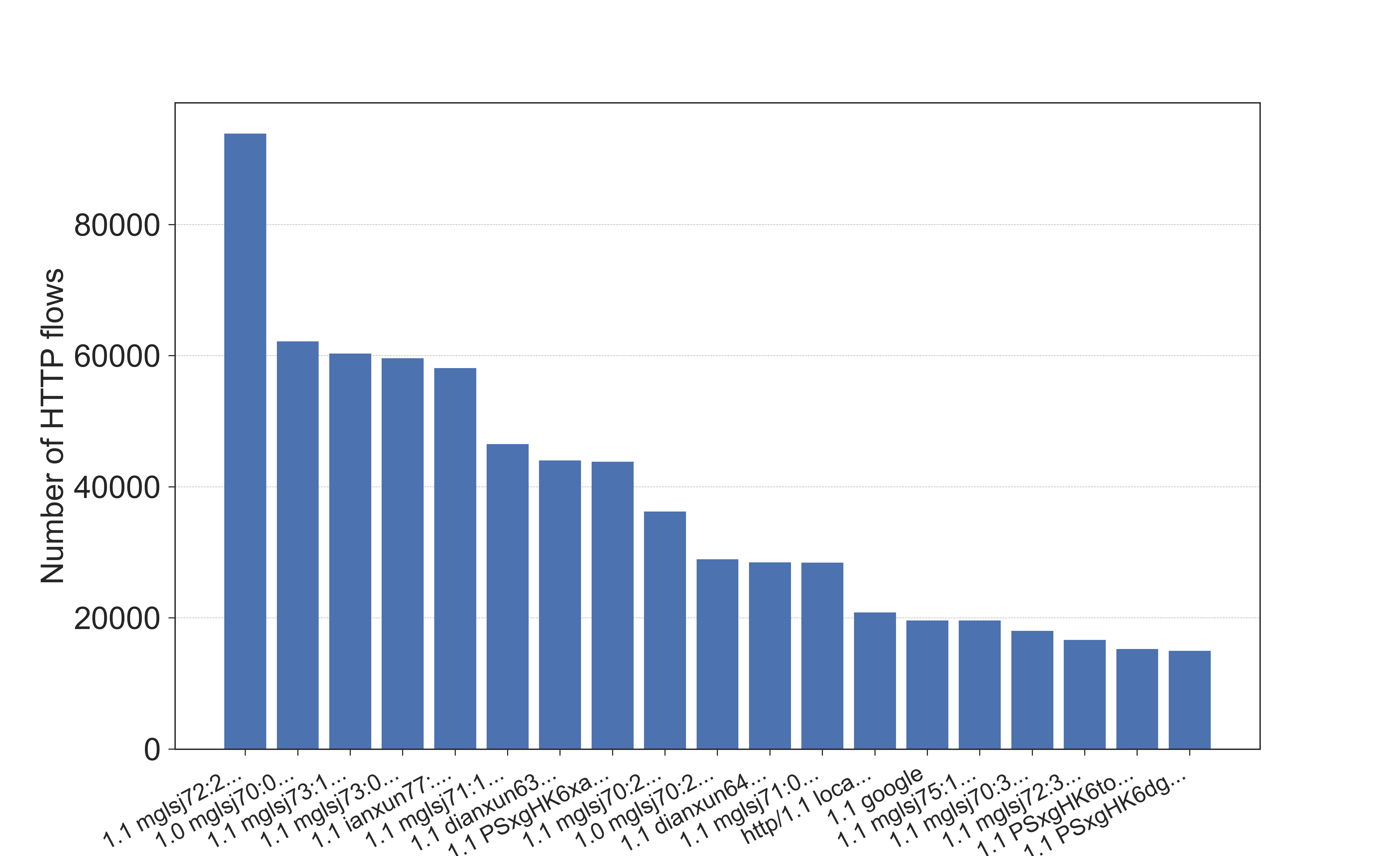}
\end{minipage}
}
\centering
\caption{Content type, response line and via fields in HTTP flows.}
\label{fig9}
\end{figure*}

For the high usage of HTTP traffic, we explore the deployment of server software on IPv6 websites. Figure \ref{fig6} reveals the difference between our active scanning and passive measurement dataset. In the Alexa top 10,000 websites, it is obvious that \textit{Cloudflare-nginx} dominates the number of famous websites. \textit{Nginx} and \textit{gws} rank 2 and 3. However, in the real IPv6 world, more than 80\% HTTP flows contain \textit{Microsoft IIS httpd} in the server field. It shows a huge shift from \textit{Cloudflare-nginx} (active Alexa 10,000) to \textit{Microsoft IIS httpd} (passively collected data). We consider the high appearance frequency of \textit{Microsoft} in HTTP host like the top 3 hosts we list in Table \ref{table3} and it causes this situation. Currently, considerable HTTP traffic flows to \textit{Microsoft} which usually have a frequent Interaction with Chinese IPv6 users. It probably caused by the massive Windows operation system due to the \verb|*.windowsupdate.com| host which is ranked 1 in Table \ref{table3}. In addition,  the server named \textit{cdn cache server 2.0} occupies a small part in the traffic. This tag belongs to \textit{Wangsu Science and Technology Co., Ltd}, which is one of the famous CDN suppliers in China. The result shows that most websites use this CDN service and overall it can be seen CDN supplier and server software are dominated by the specific individual.

%

\subsection{Protocols in IPv6 Networks}

IPv6 networks contain considerable protocols and the top 10 protocols in our measurement are HTTP, DNS, SSL/TLS, BGP, NTP, IMAPS, FTP, BOOTPS, SMTP, and NICNAEM. Among them,  HTTP, SSL/TLS, and DNS contribute most of the traffic. For these top 3 protocols, we measured the detail of the specific traffic in IPv6 networks as follows. 

\paragraph{DNS protocol}
In our measurement, we used A and AAAA records of DNS responses to discover dual-stack domains in the traffic and Verisign TLD Zone Files \cite{b36}. Figure \ref{fig7} indicates that 545,886 domains only response A record in IPv6 flows while the IPv6-only number is merely 10,744 and 35,394 domains response both A and AAAA records. In recent networks, the access frequency of dual-stack servers has remained only a small gap compared to IPv4-only servers. However, the popularity of IPv6-only server is far from the other two. It is also an indirect proof that IPv6-only websites account for a small part in the networks and they are frequently ignored by users even they have connected with IPv6 networks. However, during the accelerating IPv6 deployment, the percentage of unique IPv6-only domains appearing in the traffic seems to be 22 fold of the normal IPv6 status (2.44\% and 0.11\% respectively in accelerating deployment traffic and global .com and .net domains). The percentage of total AAAA record domains is 7.32\%, which is a 1.7-fold increase in the normal status provided by Verisign. The accelerating deployment indeed increases the user request to the IPv6-enable websites.

\paragraph{SSL/TLS protocol} We describe the details of the SSL/TLS traffic in Figure \ref{fig8}. First, Figure 9(a) shows the SSL/TLS version in IPv6 traffic, TLS1.2 occupies the most proportion except the version of the record layer in the client. TLS1.0 is still common in the record layer in the client. A few clients in IPv6 networks use SSLv3 for communication which has been forbidden by many services. In IPv6 SSL/TLS traffic, SSL/TLS certificate is one aspect we concerned. In Figure 9(b), the obviously most certificate Issuer CN is \textit{COMODO ECC Domain Validation Secure Server CA 2}. It reveals its dominance in current IPv6 issuers. In addition, Figure 9(c) observes the validity of the certificates in our dataset. It indicates that the certificate lifetime is concentrated on the interval between 2015 and 2021 in IPv6 networks. The most certificates start or end around 2018 and contain nearly one-year lifetime. Moreover, the lifetime of valid certificates tends to become shorter both in IPv4 and IPv6 network. This phenomenon is caused by the recent limitation of certificates lifetime \cite{b39}. However, this distribution of IPv4 is obviously more concentrated (85.2\% less than one-year certificate lifetime in IPv4 vs. 73.5\% in IPv6). Due to our small dataset of IPv4, it still worth larger measurement work in the future.


\paragraph{HTTP protocol}For the measurement of HTTP traffic, Figure 10(a) shows the top 10 content types in IPv4 and IPv6 HTTP flows. It proves that stream type is obviously the top content type in HTTP and the rest top 10 file types include \verb|cab|, \verb|html|, \verb|jpeg| and other content types in IPv6 networks. Application, text, and image take more proportion than other content types. In addition, we collected the response line in HTTP traffic in Figure 10(b) and found that 206 Partial Content in HTTP/1.0 or HTTP/1.1 is nearly 3 times than 200 OK. It means that IPv6 is usually used to transfer large data. These results are quite different from IPv4. Figure 10(c) explores the proxy between the user agent and the server by \verb|via| field in the header. According to our observation, over 1 million flows indicate the presence of intermediate protocols and recipients. The proxy code-named \verb|mglsj*| occupies the most proportion and 18 of the top 20 contain the \verb|cdn cache server v2.0| tag of the CDN supplier company we mention in Section \ref{test}. It proves the high usage of this CDN service during the IPv6 deployment in China.



\section{Security Issues and Common Nature During Accelerating IPv6 Networks Deployment}\label{P12}

After grasping the accelerating IPv6 deployment networks in China, we track the security issues and common nature behind the accelerating status during the accelerating deployment of IPv6 networks in this section.

\begin{figure}
\begin{center}
\scalebox{0.18}{
\centerline{\includegraphics{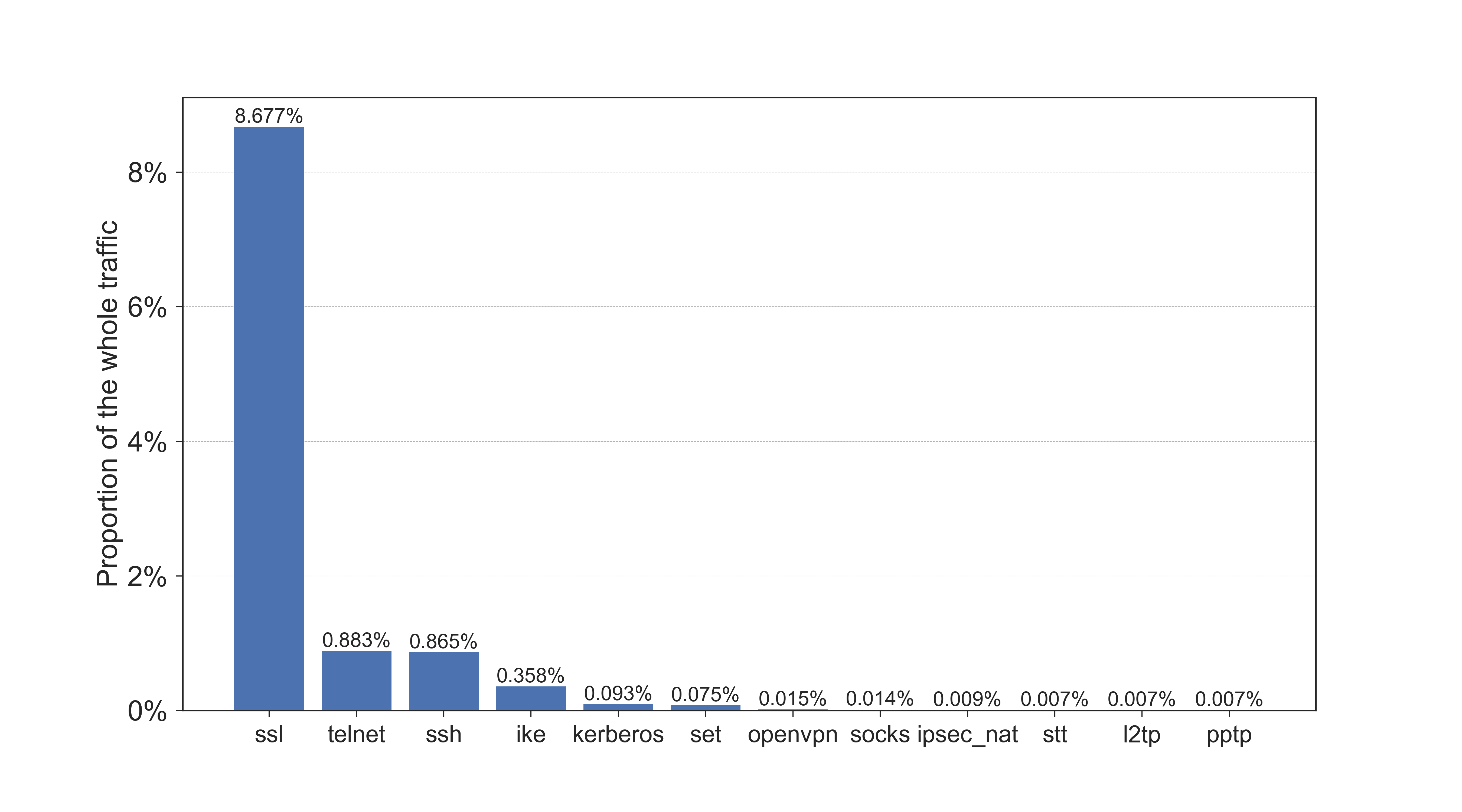}}
}
\caption{Security protocols application in IPv6 flows.}
\label{fig10}
\end{center}
\end{figure}


\begin{table}
  \caption{Access Methods in China and the World}
  \begin{center}
  \label{table5}
  \begin{tabular}{|c|c|c|c|c|}
    \hline
    \multirow{2}*{\textbf{Access Methods}}&\multicolumn{2}{|c|}{\textbf{Addresses}} &\multicolumn{2}{|c|}{\textbf{Percentage}}\\
    \cline{2-5}
    &China&World&China&World\\
    \hline
    \hline
    Teredo&490,264& 1,369,258&30.33\% &30.34\%\\
    \hline
    6to4&33,343& 93,526&2.06\% &2.07\%\\
    \hline
    6over4&13,585 &38,078&0.84\% &0.85\%\\
    \hline
    ISATAP&557 & 1,491&0.04\% & 0.03\%\\
    \hline
    Native or Others&1,078,663& 3,010,587&66.73\% & 66.71\%\\
    \hline
    \hline
    \textbf{Totally}&1,616,412 &4,512,940&100\%&100\%\\
    \hline
\end{tabular}
\end{center}
\end{table}

\begin{table*}
  \caption{Protocol Type During World IPv6 Day, World IPv6 Launch Day, and IPv6 Accelerating deployment in China}
  \begin{center}
  \label{table6}
  \begin{tabular}{|c|c|c|c|c|c||c|c|c|c|c||c|c|c|c|c|}
    \hline
    \multirow{4}*{\textbf{Period}}&\multicolumn{5}{|c||}{\multirow{2}*{\textbf{\shortstack{ World IPv6 Day\\2011-06-07 23:45 to 2011-06-09 00:45}}}}&\multicolumn{5}{|c||}{\multirow{2}*{\textbf{\shortstack{World IPv6 Launch Day\\2012-06-05 23:45 to 2012-06-07 00:45}}}}&\multicolumn{5}{|c|}{\multirow{2}*{\textbf{\shortstack{IPv6 Accelerating deployment in China\\2018-04-01 00:00 to 2018-06-30 23:59}}}}\\
    &\multicolumn{5}{|c||}{}&\multicolumn{5}{|c||}{~}&\multicolumn{5}{|c|}{~}\\
    \cline{2-16}
    &\multicolumn{2}{|c|}{\textbf{\% Protocol Type}}&\multicolumn{3}{|c||}{\textbf{\% in TCP/UDP}}&\multicolumn{2}{|c|}{\textbf{\% Protocol Type}}&\multicolumn{3}{|c||}{\textbf{\% in TCP/UDP}}&\multicolumn{2}{|c|}{\textbf{\% Protocol Type}}&\multicolumn{3}{|c|}{\textbf{\% in TCP/UDP}}\\
    \hline
    &TCP&UDP&DNS&SSL&HTTP&TCP&UDP&DNS&SSL&HTTP&TCP&UDP&DNS&SSL&HTTP\\
    \hline
    Start$^{\mathrm{*}}$&55.61&10.39&78.97&0.24&75.68&90.54&1.36&70.59&1.35&30.70&75.40&20.71&56.65&12.81&40.44\\
    Mid$^{\mathrm{*}}$&52.22&13.92&81.13&0.57&80.98&90.48&2.60&77.33&3.49&53.70&77.75&18.40&58.89&11.21&50.21\\
    End$^{\mathrm{*}}$&49.87&16.45&70.67&0.68&76.98&89.62&4.56&78.44&4.01&57.38&81.51&14.63&70.17&9.70&70.44\\
    \hline
    Overall&53.71&14.32&80.96&0.47&79.86&90.43&2.67&77.94&3.21&54.96&79.05&16.95&63.04&10.71&58.22\\
    \hline
    \multicolumn{16}{l}{$^{\mathrm{*}}$The start, mid and end divided by CAIDA refer to 1 hour respectively start from 7 June 23:45 UTC, 8 June 13:00 UTC, 8 June 23:45 UTC during World}\\
    \multicolumn{16}{l}{IPv6 Day 2011 and from 5 June 23:45 UTC, 6 June 13:00 UTC and 6 June 23:45 UTC during IPv6 Launch Day 2012, which is compared with our traffic}\\
    \multicolumn{16}{l}{data in April, May and June 2018.}
\end{tabular}
\end{center}
\end{table*}

\subsection{Security Issues}

\paragraph{User Privacy Threat}In order to grasp the security situation of the complex IPv6 networks at present, we performed the experiments related to security issues in our measurement. Figure \ref{fig10} shows the security protocols applied in the IPv6 networks. 8.677\% of the traffic belongs to SSL/TLS flows while the other security-related protocols occupy less than 1\%. In addition, IPv6 enables mechanisms with extended headers to add options header in the packets, which can be used for many security strategies. However, in our measurement, it was found that only 0.0006\% of the IPv6 traffic using the extended headers whose next header field is not TCP, UDP, or ICMPv6. Finally, we used the information entropy algorithm to detect encrypted traffic in the IPv6 world. The information entropy equation is shown as follow:
\begin{equation}
H(x) = -\sum_{i=1}^np(x_i)logp(x_i)
\end{equation}
We obtain entropy value $H(x)$ by calculating the frequency of the characters $x_i$ appear in the payload of each packet $p(x_i)$ and find that the payload entropy of the encrypted packet is usually in a confidence interval greater than 6.5. Through this method, it indicates that the encrypted traffic occupies 30.4\% in the whole IPv6 traffic. This seems very low in comparison to IPv4 networks. Due to the lack of Chinese IPv6 websites construction exposed in Table \ref{table4}, the service deployment including security service still keeps a low level. User privacy issues are about to face great challenges.


\paragraph{Inappropriate Access Methods}During the large-scale deployment of IPv6 networks period, complex networks exist various access methods. IPv4 and IPv6 networks have coexisted for a long time, the use degree of IPv4/IPv6 transition technology is necessary to be cognition. In the process of passive measurement, we explore the methods Chinese and global users access IPv6. Table \ref{table5} reveals the specific types of IPv6 connections in the network.  Except for native or unknown IPv6 access methods, using Teredo technology is more common than other known IPv6 access methods. In the national IPv6 network, the distribution of access methods of Chinese hosts is similar to the world. More than 30\% of users access the IPv6 networks by using the IPv4/IPv6 transition and the IPv6 development will face a severe problem in the future. Complex network structure will bring long-term risks to the IPv6 network.

\subsection{Common Nature}

World IPv6 Day and World IPv6 Launch were announced to motivate organizations across the industry to prepare their services for IPv6. To record the traffic, CAIDA \cite{b32} provides the anonymized Internet traces on IPv6 Day and IPv6 Launch Day dataset. This dataset is also collected during the accelerating deployment event which is similar to our measurement. We combine the past and current traffic in our work to explore the common nature during the IPv6 accelerating deployment. Table \ref{table6} summaries the detail of the work. 

In the past IPv6 networks in 2011 and 2012, UDP traffic seems to have a continuously increase during the period of accelerating deployment, while today in our traffic TCP is the increasing one. In addition, we discover that TCP traffic is at the leading position in the IPv6 networks. According to the measurement, we notice that DNS and HTTP traffic has a sharp increase in the all 3 datasets (except World IPv6 Day which is not keeping to the end) and HTTP even has a 74.2\% growth rate from 40.44\% to 70.44\% in TCP traffic during the major event today. It can be one reason for the rising TCP traffic. DNS traffic also has a share of up from 56.65\% to 70.17\% in UDP today. In the three specific days of major IPv6 events, all the three application layer protocols have a small growth time except SSL/TLS during Chinese IPv6 large event today. Even the SSL/TLS traffic volume has increased for three months but it goes backward for the ratio. We consider that SSL/TLS deployment may not satisfy the great increase of IPv6 traffic and future networks need to improve these issues for the security of privacy in the IPv6 traffic.

\section{Discussion}\label{P5}

\subsection{Current Accelerating Status in China}
At present, China is accelerating the large-scale deployment of IPv6. Active addresses and traffic are increasing dramatically. The IPv6 address prefixes are also increasing at a rate of 10.1 per day and the traffic volume has a 10-fold growth. This growth trend will not stop and so far there is no sign of a recession. However, practices indicate that the current accelerating status is unbalanced and unstable in all four metrics we measured. It also exposes many issues including insufficient encrypted application usage, low content encryption rates, and excessive IPv4/IPv6 transition usage. The improvement of network performance conflicts with the challenge of network security and stability. We conclude that this impact will be long-term and will not be resolved until future progress in the network.

\subsection{Future Accelerating IPv6 Deployment}
Accelerating network deployments usually have an unbalanced and unstable network status like China. In order to solve the situation of network imbalance, in addition to increasing the network address and network access environment, we believe that IPv6 websites and network services construction is the key to the accelerating stage. In addition, we found that there are often security issues when deploying IPv6 on a large scale. When accelerating IPv6 networks deployment are been implementing, HTTP traffic will increase significantly but SSL/TLS usage tends to grow slowly, which is also closely related to the security service deployment of the website. Therefore, security service deployment is the next pending requirement for accelerating IPv6 deployment. When the network hosts and network services are balanced, we will see a better IPv6 future than today's beginning.

\section*{Acknowledgment}

This work is supported by The National Natural Science Foundation of China (No.U1636217 and No.61702501) and The National Key R\&D Program of China (No.2016YFB0801200 and No.2016QY05X1000) and The Key research and Development Program for Guangdong Province under grant No.2019B010137003. Gaopeng Gou is the corresponding author. The authors would like to also thank MAXMIND, Arbor Networks, Route Views, Google, Verisign, Amazon, and CAIDA  for making data available.

\vspace{12pt}
%
\end{document}